\documentstyle[12pt,epsf]{article}

\catcode`\@=11
\def\vereq#1#2{\lower3pt\vbox{\baselineskip1.5pt \lineskip1.5pt
\ialign{$\m@th#1\hfill##\hfil$\crcr#2\crcr\sim\crcr}}}
\catcode`\@=12
\def\href#1#2{{#2}}

\begin{document}
\begin{titlepage}
\begin{center}
\today     \hfill    
 \hfill MN-HEP-1527, TPI-01/97\\
 \hfill KEK-TH-509,  TIT/HEP-359\\
 \hfill LBNL-39902, UCB-PTH-97/05\\
 \hfill hep-ph/9702202


{\large \bf Upper Bound on $R_b$ in Two Higgs Doublet Model\\
from Lepton Universality}\footnote{The work of J.H. was supported by
the DOE under contract DE-FG02-94ER40823.  
The work of H.M. was supported in part by the DOE under contract
DE-AC03-76SF00098, and in part by the NSF under grant PHY-95-14797.}  
\vskip 0.3in

Junji Hisano,$^1$  Shingo Kiyoura,$^{2,3}$ and
Hitoshi Murayama$^{4,5}$\footnote{Alfred P. Sloan Fellow.}

\vskip 0.3in
{\small 
$^1$ {\em School of Physics and Astronomy,
     University of Minnesota\\
     116 Church Street S.E.\\
     Minneapolis, MN55455, USA}\\ 
$^2$ {\em Department of Physics, Tokyo Institute of Technology\\
Oh-okayama 2-12-1, Megro-ku, Tokyo, 152, Japan}\\
$^3$ {\em National Laboratory of High Energy Physics\\
     1-1 Oho, Tsukuba-shi,
      Ibaraki-ken, 305, Japan}\\ 
$^4$ {\em Department of Physics,
     University of California\\
     Berkeley, California 94720, USA}\\
$^5$ {\em Theoretical Physics Group\\
     Ernest Orlando Lawrence Berkeley National Laboratory\\
     University of California,
     Berkeley, California 94720, USA}
}        
\end{center}



\begin{abstract}
It has been known that $R_b$ can be enhanced in the two Higgs doublet 
model if $\tan \beta$ is large.  We point out that a similar 
enhancement in $\Gamma (Z \rightarrow \tau^+ \tau^-)$ is large enough 
to place a constraint on such a possibility.  We obtain a 95\% CL 
upper bound $\Delta R_b/R_{b} < 0.73\%$ in this model for the 
$\overline{\rm MS}$ mass $m_{b} (m_{Z}) = 3.0$~GeV.  The 1996 world
average is $\Delta R_{b}/R_{b}$ = $0.97\% \pm 0.51\%$.  We used the 
$m_{b}(m_{Z})$ to determine the bottom Yukawa coupling instead of 
$m_b(m_{b})$ unlike in previous analyses, and also an improved experimental 
test of the lepton universality in $Z$ decay, which made our results 
qualitatively different.
\end{abstract}

\vfill

\end{titlepage}

\renewcommand{\thepage}{\arabic{page}}
\setcounter{page}{1}

The Standard Model (SM) has been tested at an impressive accuracy by 
recent collider experiments such as LEP, SLC, and Tevatron.  Currently 
no serious conflict between the data and the SM is reported.  However, 
the $Z \rightarrow b\bar{b}$ branching fraction has been higher than 
the SM value at a few standard deviation level.  The value of $R_{b} 
\equiv \Gamma(Z \rightarrow b\bar{b}) / \Gamma( Z \rightarrow 
\mbox{hadrons})$ reported in summer 1995 \cite{Beijing} was actually 
more than three standard deviations higher from the SM prediction and 
stimulated many theoretical and experimental efforts.  The most recent 
measurements from SLC ($0.2149\pm 0.0032)$ \cite{SLC-Rb}, ALEPH 
$(0.2158\pm 0.0009)$ \cite{ALEPH-Rb}, DELPHI 
$(0.2176\pm0.0028\pm0.0027)$ \cite{DELPHI-Rb}, and OPAL 
$(0.2175\pm0.0014\pm0.0017)$ \cite{OPAL-rb}, however, are consistent 
with the SM ($R_b^{SM}=0.2157$ for $m_t=175$ GeV), and as a result, 
the 1996 world average value combining all the old and new data has 
come closer to the SM prediction $(R_b=0.2178\pm0.0011)$ 
\cite{alllep}.  Since older and newer measurements have different 
systematics, it is not clear at this stage whether it is appropriate 
to discard older measurements from the world average.  In fact, DELPHI 
suggested to combine their older and newer measurements, while ALEPH 
did not.  The central 
value may further evolve as newer methods are developed and the 
experimental inputs are updated.  In view of this situation, it is 
premature to judge what the final outcome would be.  Therefore, it is 
useful to investigate the consequence of various models on $R_{b}$ 
in the light of other experimental constraints.

It has been argued that the 1996 $R_{b}$ value is actually more favored
by various new physics scenarios than the 1995 one
\cite{Kane-Warsaw,Pokorski-Warsaw,Langacker-APS}.  The 1995 average
was hard to be explained by one-loop corrections to the $Zb\bar{b}$
vertex due to new particles.  The 1996 average, on the other hand, is
within the variation of $R_{b}$ values predicted in the many new physics
scenarios.  This arises a renewed interest to check the consistency
of various scenarios of high $R_{b}$ with other existing
experimental constraints.

Currently there are two popular models which may lead to $R_{b}$ 
values higher than that in the SM. One is the Minimal Supersymmetric 
Standard Model (MSSM) in the small $\tan\beta$ region, where the loop 
of a scalar top (mostly right-handed one) and a chargino (mostly 
higgsino-like one) can enhance $R_{b}$ which attracted many 
discussions \cite{MSSM}.  The other is the loop of pseudo-scalar and 
scalar Higgs bosons in two Higgs doublet model (2HDM) which can also 
enhance $R_{b}$ if $\tan\beta$ is large.  The first scenario is now 
strongly constrained by recent limits on the chargino mass from LEP-II 
and the scalar top mass from D0.  Still, one can accommodate a 
correction to $R_b$ as large as $\Delta R_b < 0.0017$ \cite{ELN}.  It 
was argued that the latter scenario is in a conflict with the lack of 
an enhancement in the four-$b$ final states from $Z$ decay \cite{KW} 
(see also \cite{KK}).  The current experimental limit on four-$b$ or 
$\tau^+ \tau^- q\bar{q}$ final state \cite{Yukawa-process} is, 
however, not as stringent as estimated in \cite{KW}, and there 
still remains a possibility that this scenario may enhance the $R_{b}$ 
at a desirable level.  It is the purpose of this letter to investigate 
whether this scenario is consistent with other existing constraints.

The 2HDM has several motivations.  First of all, it is the simplest
extension of the minimal standard model which has to be confronted by
experiments.  Second, the electroweak baryogenesis requires an extension
of the minimal standard model to incorporate a CP-violation in the Higgs
sector.  The 2HDM can naturally have CP-violating phases in its potential,
and it is argued that it can create the value of cosmic baryon asymmetry
as required by nucleosynthesis (for a review, see \cite{baryogenesis}).  
Third, the 2HDM may be a part of the MSSM.
A general 2HDM has a potential problem of flavor-changing neutral
currents, which can be naturally avoided by either of the following two
ways.  The Type-I 2HDM lets only one of the Higgs doublets couple to
quarks and leptons and hence the coupling matrix of the Higgs bosons can
be simultaneously diagonalized as the mass matrix of quarks.  There is
no flavor-changing vertex of the Higgs bosons.  In this case, however, a
large $\tan\beta$ does not enhance the Higgs coupling to the $b$-quark,
and hence there is no interesting large contribution to $R_b$.  On the
other hand, the Type-II 2HDM lets one of the Higgs bosons couple to the
up-quarks while the other to the down-quarks and a large $\tan\beta$ can
enhance the coupling to the $b$-quark.  In principle, either one of them
can couple to the leptons.  It is probably natural, however, to assume
that the one which couples to the down-quarks also couples to the
leptons because they share the same weak isospin $I_3 = -1/2$.  It is
indeed the case, for instance, in the MSSM.  Then it is a natural question
to ask whether a large $\tan\beta$ affects the phenomenology of the
lepton sector.

We find that there is a strong correlation between $\Gamma(Z 
\rightarrow \tau^+\tau^-) / \Gamma(Z \rightarrow e^+ e^-)$ and $R_b$ 
in the Type-II 2HDM. From the observed lepton universality in 
$Z$ decay, we find that $R_b$ cannot deviate from the SM 
prediction by more than $0.73\%$ at 95\% confidence level
almost independently from $\tan\beta$.  This upper bound is in 
a conflict with the current central value of $R_b$.  

The consequence of the 2HDM on the $Zb\bar{b}$ and $Z\tau^+\tau^-$ 
vertices was studied already some time ago \cite{2HDM}.  It was 
concluded that the correction to $Z\tau^+\tau^-$ vertex was too small 
to be observed compared to its size to $Zb\bar{b}$ vertex.  We point 
out that two important changes should be made to this conclusion, 
however.  The first is the improved accuracy in the experimental test 
of the lepton universality in $Z$ decay, and the second is the running 
effect of the $b$-quark Yukawa coupling between $m_{b}$ and $m_{Z}$ 
which was not taken into account.  A combination of a high accuracy in 
$Y_\tau$ (defined below) and the running effect of $m_b$ can make the
$Z\tau^+\tau^-$ vertex much more sensitive than previously thought.

The lepton universality in $Z$ decay is now tested at an extremely 
high accuracy.  For later discussion, we parameterize the possible 
violation of the lepton universality by the following double ratio, %
\begin{equation}
Y_{\tau} = \frac{\Gamma(Z\rightarrow \tau^{+}\tau^{-})/
                \Gamma(Z\rightarrow \tau^{+}\tau^{-})_{\rm SM}}
                {\Gamma(Z\rightarrow l^{+} l^{-})/
                \Gamma(Z\rightarrow l^{+} l^{-})_{\rm SM}} .
\end{equation}
Here, $l$ refers to either $e$ or $\mu$ assuming the universality 
among them, and the subscript SM to the standard model values.  The 
advantage of using $Y_{\tau}$ is that many uncertainties cancel in the 
double ratio.  Experimentally, the uncertainties in the luminosity 
measurement and overall width measurement nearly cancel between $\tau$ 
and $l$.  Theoretically, the top quark and Higgs boson masses enter 
the predictions of partial widths through oblique corrections, but they 
cancel in the ratio.  The ratio of the SM values $\Gamma(Z\rightarrow 
\tau^{+}\tau^{-})_{\rm SM}/ \Gamma(Z\rightarrow l^{+} l^{-})_{\rm SM} 
= 0.9977$ is determined mostly by the kinematic factor $\beta_\tau^3 = 
(1-4 m_{\tau}^2/m_Z^2)^{3/2}$ to a very good approximation because of 
the axial-coupling dominance in lepton couplings to the $Z$ boson.  
Therefore, there is little theoretical ambiguity in the predicted 
ratio in the SM. We hence find that $Y_{\tau}$ is the most useful 
quantity for our purpose.

We derived the experimental value of $Y_{\tau}$ from the $Z$ line 
shape and lepton forward-backward asymmetries reported by the LEP 
Electroweak Working Group \cite{alllep}.  They quote measured values 
of $m_Z$, $\Gamma_Z$, $\sigma_{\rm h}^0$, $R_e$, $R_{\mu}$, 
$R_{\tau}$, $A_{\rm FB}^{0,e}$, $A_{\rm FB}^{0,\mu}$, $A_{\rm 
FB}^{0,\tau}$ and their errors and correlation including new data from 
the 1995 energy scan.  Here, $R_l \equiv \Gamma(Z\rightarrow 
\mbox{hadrons})/\Gamma(Z\rightarrow l^+ l^-)$.  Assuming the lepton 
universality for the first two generations, $R_l \equiv R_e = R_\mu$, 
$A_{\rm FB}^{0,l} \equiv A_{\rm FB}^{0,e} = A_{\rm FB}^{0,\mu}$, and 
using $\chi^2$ fit to the seven remaining parameters, we determine the 
ratio
\begin{equation}
\frac{R_l}{R_\tau} = 
        \frac{\Gamma(Z\rightarrow \tau^+ \tau^-)}{\Gamma(Z\rightarrow l^+l^-)} 
        = 0.99850 \pm 0.0030.  
\end{equation}
The correlations among the seven parameters are fully taken into account. 
Normalizing it by the ratio in the standard model, we obtain
\begin{equation}
Y_{\tau} =1.0008\pm 0.0030.
\end{equation}
The 95\% confidence level upper bound\footnote{The confidence level of
the upper bound is determined by the one-sided Gaussian distribution.} 
is $Y_{\tau} < 1.0057$.  Note 
that Ref.~\cite{alllep} quotes a somewhat weaker constraint on the 
ratio of electron and $\tau$ couplings $g_{V\tau}/g_{Ve}=0.959\pm 0.046$
and $g_{A\tau}/g_{Ae}=1.0000\pm 0.0019$, which correspond to 
$0.4\%$ error in $Y_\tau$.  However, the ratios do not
assume $e$-$\mu$ universality and they tried to separate axial and
vector couplings which are not necessary for our purpose.

\begin{figure}
\centerline{\epsfxsize=0.5\textwidth \epsfbox{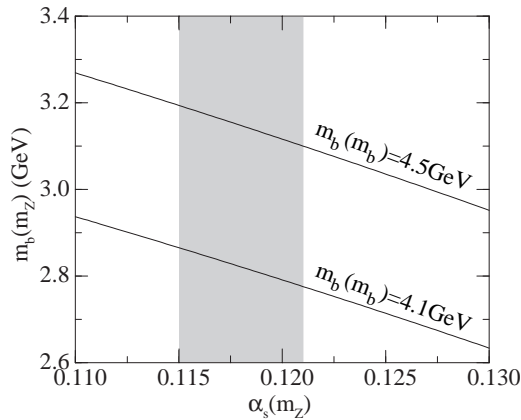}}
\caption
{ Dependence of $b$-quark mass at $m_Z$ ($m_b(m_Z)$) on 
$\alpha_s(m_Z)$.  Two solid lines correspond to $m_{b}(m_{b})$=4.1 GeV and 
4.5 GeV, respectively.  The shaded region represents the PDG average
$\alpha_{s}(m_{Z}) = 0.118 \pm 0.003$.  }
\label{fig:mb}
\end{figure}

Second, we include the effect of the running of $b$-quark Yukawa 
coupling between $m_{b}$ and $m_{Z}$ scales which was not taken into 
account in the previous analysis \cite{2HDM}.  We take the following 
procedure.  We first take the value of $\overline{m}_{b} \equiv 
m_{b}(m_{b}) = 4.1$--$4.5$~GeV in the $\overline{\rm MS}$ scheme as 
summarized by the Particle Data Group \cite{PDG96}.  To run the 
$\overline{\rm MS}$ mass from $m_{b}$ to $m_{Z}$, we employ the 
renormalization group equation at two-loop level \cite{2loopmb}.  The 
numerical values of $m_{b}(m_{Z})$ are shown in Fig.~\ref{fig:mb}.  The shaded 
region is the world average $\alpha_{s}(m_{Z}) = 0.118 \pm 0.003$ 
\cite{PDG96}.\footnote{%
Note that this average does not include the total $Z$ hadronic width 
which is consistent with our spirit to allow $R_{b}$ to deviate from 
the SM prediction.  } %
In this letter we take center values of $\overline{m}_{b}$ and 
$\alpha_s(m_Z)$, which correspond to  $m_{b}(m_{Z})=3.0$GeV.  We will 
discuss later how the results change for different choices of 
$\overline{m}_{b}$ and $\alpha_{s} (m_{Z})$.

We do not go into the discussion of the MSSM Higgs sector in our letter.
This is partly to simplify the analysis without many additional
parameters in the MSSM.  It is however mainly because a large
contribution to $R_b$ requires a large $\tan\beta$ with a light Higgs
multiplets, which in turn implies a light charged Higgs boson in the
context of the MSSM.  Such a light charged Higgs boson is already
strongly constrained by the CLEO measurement of the $b\rightarrow
s\gamma$ rate.  One needs to rely on a cancelation between the charged
Higgs diagram (which always adds up with the standard model
contribution) and the chargino loop which is also uncomfortably 
large.  A light charged Higgs boson is further constrained also by $B \rightarrow 
\tau \nu X$ \cite{Yuval} and $\tau \rightarrow \mu \nu \nu$ \cite{taudecay}.  
Even though there are viable regions in the parameter space \cite{MSSM},
we do not pursue this direction further in this letter.\footnote{There
is a possibility that the scalar tau loop may partially cancel the
enhancement in $\Gamma(Z \rightarrow \tau^{+} \tau^{-})$ while keeping
$R_{b}$ large in a limited region of the parameter space.  It is an
interesting question whether such a cancelation is possible within the
parameter space which sufficiently suppresses $b\rightarrow s\gamma$
\cite{Kiyoura}.}

On the other hand, a general Type-II 2HDM allows a heavy charged Higgs 
boson naturally consistent with the $b\rightarrow s\gamma$ constraint, 
while having neutral Higgs bosons as light as 50~GeV\footnote{The charged Higgs boson mass must be heavier than about
244~GeV due to the constraints from $b\rightarrow s\gamma$ \cite{CLEO}.  
On the other hand, the electroweak $\rho$-parameter restricts the mass
splitting among the Higgs bosons.  For light $h$ and $A$, we estimate
the upper bound on the charged Higgs boson mass to be $m_{H^\pm}\sim 200$~GeV at 95\% CL using the precision electroweak data given
in \cite{alllep}.  Therefore, the 2HDM with light $h$ and $A$ is either
only marginally consistent with these constraints, or requires some new
physics to be consistent with the electroweak precision data.  We simply
assume in this letter that such new physics enters only oblique
corrections and does not modify $Z\bar{b}$ or $Z\tau^+ \tau^-$ vertices. 
We thank S.Kanemura for discussions on this point.}. The 
phenomenological viability of such a light Higgs spectrum was recently 
also stressed in \cite{Krawczyk}.  This is the parameter range of our 
interest in this letter.

The 2HDM contains five physical Higgs bosons, two neutral CP even 
states $h^0$ and $H^0$, one CP odd state $A^0$, and two charged states 
$H^+$ and $H^-$.  We decompose two Higgs doublets as
\begin{equation}
H_1 = \left(
        \begin{array}{c}
        \frac{1}{\sqrt{2}} 
        \left(v \cos\beta + \eta_1 + i \xi_1\right) \\
        H_1^-
        \end{array} \right),
\quad
H_2 = \left(
        \begin{array}{c}
        H_2^+ \\
        \frac{1}{\sqrt{2}} 
        \left(v \sin\beta + \eta_2 + i \xi_2\right)
        \end{array}
        \right).
\end{equation}
The mass eigenstates of the Higgs bosons are related 
to the weak eigenstates as follows:
\begin{eqnarray*}
h^0 &=&  - \eta_1 \sin\alpha + \eta_2 \cos \alpha, \\
H^0 &=&   \eta_1 \cos \alpha + \eta_2 \sin \alpha,
\end{eqnarray*}
for lighter ($h^0$) and heavier ($H^0$) neutral CP-even Higgs bosons, 
\begin{displaymath}
A^0 =  \xi_1 \sin\beta + \xi_2 \cos \beta,
\end{displaymath}
for the neutral CP-odd Higgs boson, and 
\begin{displaymath}
H^- =  H_1^- \sin\beta + (H_2^+)^* \cos \beta,
\end{displaymath}
for the charged Higgs boson.  At this point, we have the following free
parameters: $m_{h^0}$, $m_{H^0}$, $m_{A^0}$, $m_{H^\pm}$, $\alpha$,
$\beta$.  Below, we take the limit of the heavy charged Higgs boson
which is possible in the general 2HDM parameter space.  We do so because
the loop of the charged Higgs boson and the top quark reduces $R_{b}$
and hence the positive contribution to $R_{b}$ is maximized in this
limit.  Note also that the constraint from $b\rightarrow s\gamma$ is
naturally avoided in this limit as well.

\begin{figure}
\centerline{\epsfxsize=0.5\textwidth \epsfbox{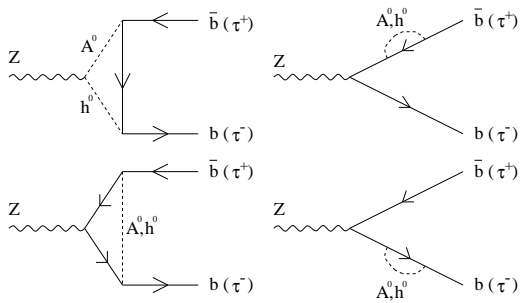}}
\caption
{Feynman diagrams with the neutral Higgs bosons in the 2HDM which 
contribute to the partial widths $Z\rightarrow b\bar{b}$ and 
$\tau^+\tau^-$ and have the $\tan^2 \beta$ enhancement.  }
\label{fig:diagrams}
\end{figure}

We evaluated the diagrams shown in Fig.~\ref{fig:diagrams}, and found an
agreement with previous calculations \cite{2HDM}.\footnote{ %
We ignored $b$-quark and $\tau$-lepton masses in loop diagrams while 
we keep their Yukawa coupling constants finite.  The vertex diagrams 
$Z\rightarrow \mbox{virtual $(Zh^0$ or $H^0)$} \rightarrow b \bar{b}$ 
are proportional to $m_b\tan\beta$ but lack the $\tan^2\beta$ 
enhancement.  For large values of $\tan\beta$ we are interested in, 
this contribution is subdominant and the omission of $m_b$ in the 
diagram is justified.  The diagrams with Nambu--Goldstone bosons in 
the $R_\xi$ gauge are the same as in the SM and hence are not enhanced 
for large $\tan\beta$.  } All the loop factors are common for $b$ and 
$\tau$.  The only differences are in the size of Yukawa coupling 
$\lambda_{f}$ and the couplings to the $Z$ boson, $g_{L}^{f} = 
-\frac{1}{2} - Q_{f} \sin^{2}\theta_{W}$ and $g_{R}^{f}=-Q_{f} 
\sin^{2}\theta_{W}$ for $f=b,\tau$.  By parameterizing the loop factors 
from the first (second) diagram as $\epsilon_{hA}$ ($\epsilon_{ff}$), 
the widths are corrected as
\begin{equation}
        \frac{\Delta \Gamma_{f}}{\Gamma_{f}}
                = \frac{\lambda_{f}^{2}}{(g_{L}^{f})^{2} + (g_{R}^{f})^{2}}
                \left[ 2 (g_{R}^{f} - g_{L}^{f}) \frac{1}{2} \epsilon_{hA} -
                        4 g_{R}^{f} g_{L}^{f} \epsilon_{ff} \right] .
\end{equation}
Both $\epsilon_{hA}$ and $\epsilon_{ff}$ turn out to be positive.  For 
both $b$ and $\tau$, $g_{R}^{f} - g_{L}^{f} = 1/2$ and hence the first 
term in the bracket is simply $\epsilon_{hA}/2$.  On the other hand, 
the second term is $0.25\epsilon_{ff}$ for $\tau$ and 
$0.13\epsilon_{ff}$ for $b$, and hence is less important for $b$.  %
In order to simplify the discussion we take $\alpha=\beta$ where the
$H^0$ is almost decoupled from leptons and down-quarks.  In this case
$h^0$ has the enhanced Yukawa coupling for large $\tan\beta$ while $H^0$
is the SM Higgs boson.  The value of $m_{H^0}$ is irrelevant
to the following discussion.\footnote{Note that there is a dependence
on the mass of the standard model Higgs boson at 0.2\% level when one
predicts partial widths theoretically.  However this dependence comes
through oblique corrections and cancels in the ratios $R_b$ and
$Y_\tau$.}  One can easily generalize the analysis for
$\alpha\neq\beta$, but the strong correlation remains the same and hence
the final conclusion as well.  The discussion then depends only on the
following three parameters: $\tan\beta$, $m_{A^0}$, and $m_{h^0}$.

\begin{figure}
\centerline{\epsfxsize=0.5\textwidth \epsfbox{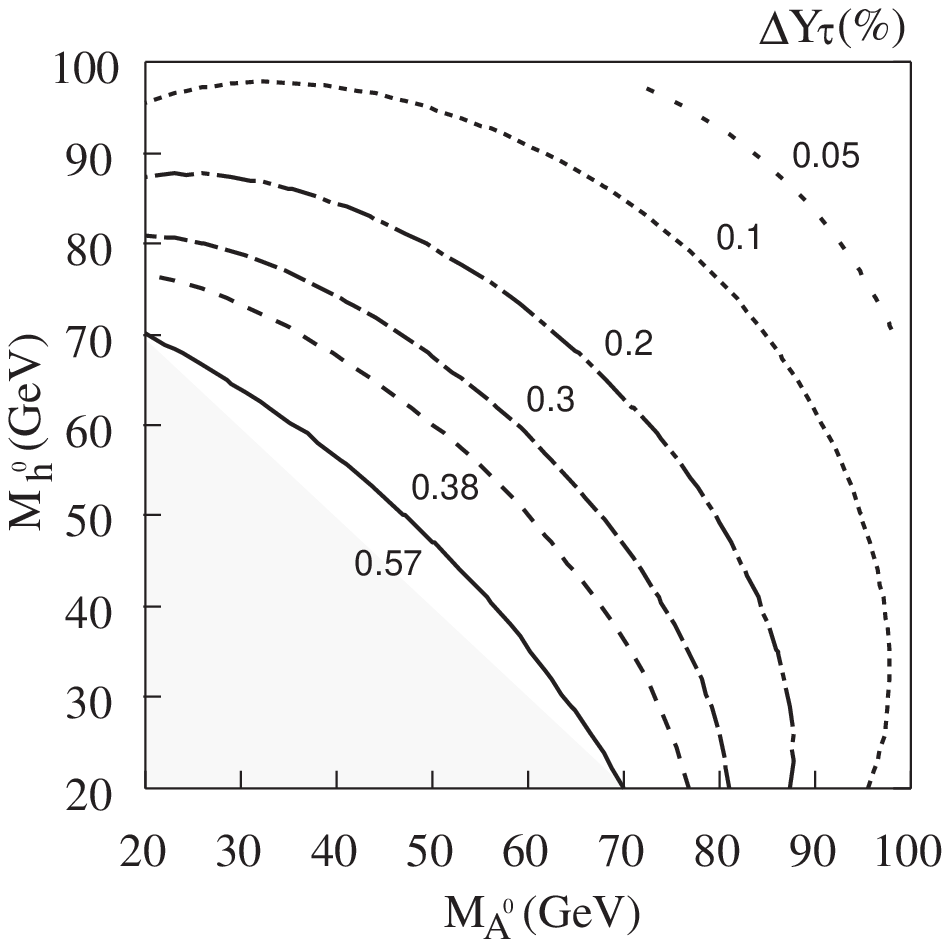}}
\caption
{A contour plot of the $\Delta Y_{\tau}/Y_{\tau}$ induced by the 
neutral Higgs bosons on the $(m_{h^0}, m_{A^0})$ plane 
with $\tan\beta=70$.  The shaded region is excluded by the negative 
search for $Z\rightarrow A^0~h^0$ ($m_{h^0} + m_{A^0} > m_Z$).  }
\label{fig:Ytau}
\centerline{\epsfxsize=0.5\textwidth \epsfbox{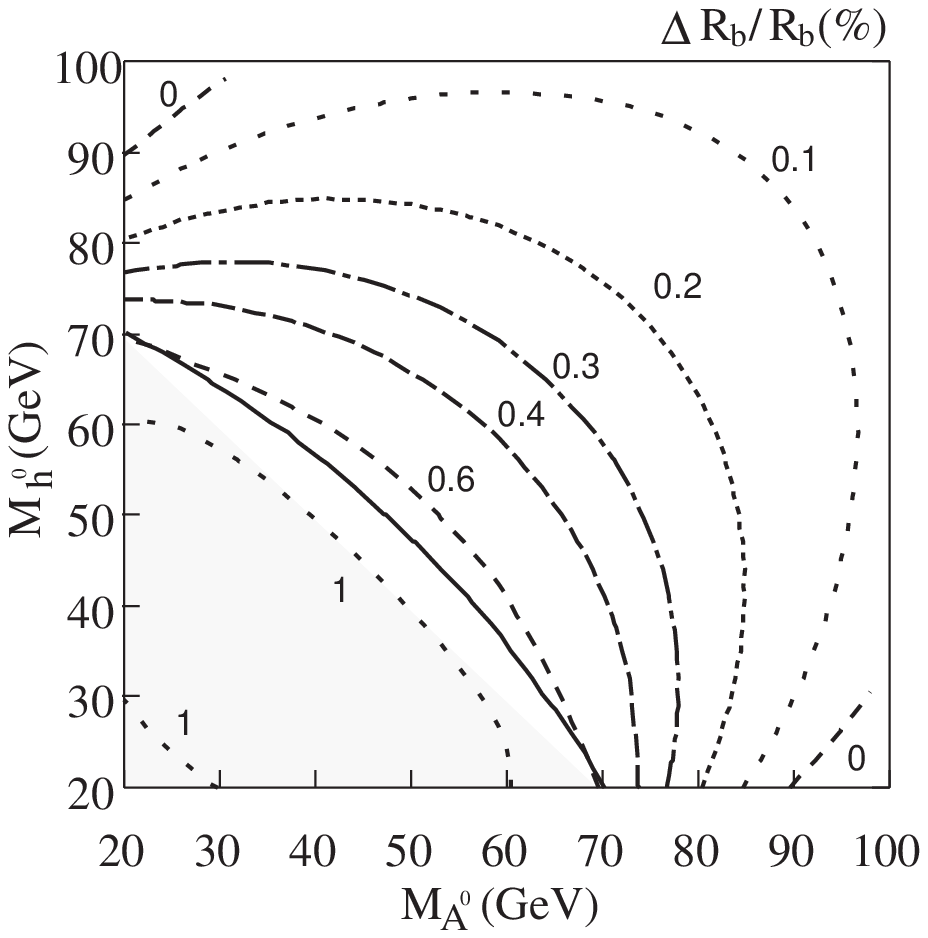}}
\caption
{
A contour plot of the $\Delta R_b/R_b$ 
induced from the neutral Higgs loops  on the $(m_{h^0}, m_{A^0})$ plane
with $\tan\beta=70$.  The shaded region is excluded by the 
negative search for $Z\rightarrow A^0~h^0$ ($m_{h^0} + m_{A^0} > m_Z$).
The solid line represents the bound from $\Delta Y_{\tau}
<0.57\%$.
}
\label{fig:Rb}
\end{figure}

The correction to $Y_{\tau}$ from the neutral Higgs loops is shown in 
Fig.~\ref{fig:Ytau}.  This is a contour plot of the $\Delta Y_{\tau}$ 
induced from the neutral Higgs loops in the $(m_{h^0}, m_{A^0})$ plane 
with $\tan\beta=70$.  The shaded region is excluded by the 
negative direct search for $Z\rightarrow A^0~h^0$ ($m_{h^0} + m_{A^0} 
> m_Z$).  The present experimental bound $\Delta Y_{\tau} < 0.0057$ is 
shown by the solid line, and is competitive with the constraint from 
the direct search.  The excluded region from $Y_{\tau}$ is wider for 
larger $\tan\beta$ because the correction is proportional to $\tan^2 
\beta$.

The correction to $R_b$ from the neutral Higgs loops is shown in 
Fig.~\ref{fig:Rb}, as a contour plot of the $\Delta R_b/R_b$ induced 
from the neutral Higgs loops with $\tan\beta=70$.  The solid line 
represents the 95\% CL upper bound $\Delta Y_{\tau}<0.57\%$.  Recall that the
central value of the 1996 world average is $\Delta R_{b}/R_{b} = 
0.97\%$.

\begin{figure}
\centerline{\epsfxsize=0.7\textwidth \epsfbox{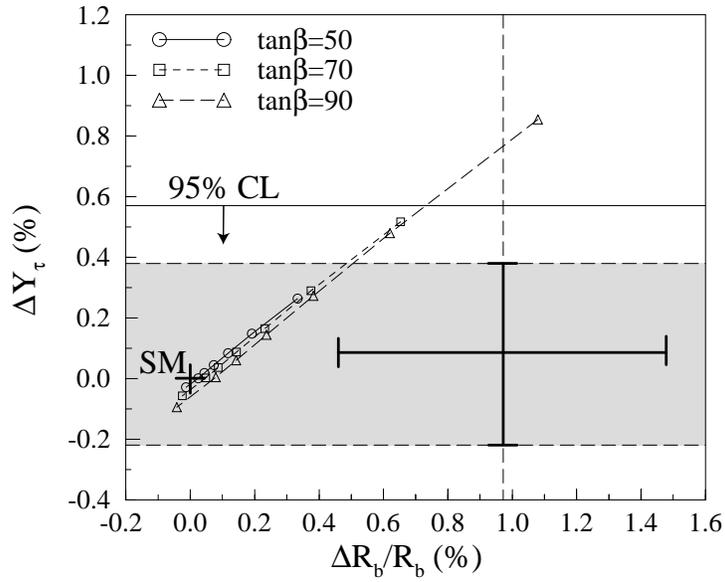}}
\caption
{Correlation between between $\Delta R_b/R_b$ and $\Delta Y_{\tau}$.  
Here we assume $m_{A^0}=m_{h^0}$ which maximizes $R_{b}$ relative to 
$Y_{\tau}$.  The curves are for $\tan\beta=50, 70, 90$.  Marks on each 
curves correspond to $m_{A^0}=m_{h^0}$= 50, 60, 70, 80, 90, 100, 150 GeV. 
Shaded region is the current experimental value of $\Delta Y_{\tau}$ 
at one standard deviation, and the solid line is the 95\% CL 
upper bound $\Delta Y_{\tau}<0.57\%$.  }
\label{fig:correlation}
\end{figure}

As can be seen from Figs.~\ref{fig:Ytau} and \ref{fig:Rb}, both 
$\Delta R_b/R_b$ and $\Delta Y_{\tau}$ become maximum in a region with 
$m_{A^0}=m_{h^0}$.  This is particularly true for $R_{b}$ because 
$\epsilon_{ff}$ is less important and $\epsilon_{hA}$ is maximized 
when $m_{A^0}=m_{h^0}$.  In order to obtain the most conservative
constraint, we take $m_{A^0}=m_{h^0}$ which maximizes $R_{b}$ while
keeping $Y_{\tau}$ small.  In Fig.~\ref{fig:correlation} we show the 
correlation between $\Delta R_b/R_b$ and $\Delta Y_{\tau}$ with 
$\tan\beta=50, 70, 90$, assuming $m_{A^0}=m_{h^0}$.  Marks in each lines 
correspond to $m_{A^0}=m_{h^0}=50, 60, 70, 80, 90, 100, 150$ GeV. 
The shaded region is the current experimental value of  
$\Delta Y_{\tau}$ at one standard deviation, and the solid line 
is the 95\% upper bound $\Delta Y_{\tau}<0.57\%$.  From this correlation, 
the current bound on $\Delta Y_{\tau}$ constrains 
$\Delta R_b/R_b< 0.73\%$.  It is interesting that the 1996 world 
average $\Delta R_{b}/R_{b} = 0.97\%$ is well beyond the 95\% CL upper bound.

Finally, we would like to discuss dependence of the upper bound on 
$R_b$ on the $b$-quark mass and $\alpha_s(m_Z)$.  The upper bound on 
$R_b$ is proportional to $m_b^2(m_Z)$, and then larger 
$\alpha_s(m_Z)$ or smaller $\overline{m}_b$ gives a more stringent constraint. 
If we take $\alpha_s(m_Z)$=0.121 (0.115) with $\overline{m}_b$=4.3GeV, 
it can be found from Fig.~1 that the upper bound on $\Delta R_b/R_b$ becomes 0.71\%
 (0.75\%). A lattice calculation on $\Upsilon$ spectroscopy gives 
$\overline{m}_b=4.0\pm0.1$GeV \cite{Davies}, and a recent QCD sum rule analysis 
using more recent data of the electronic partial 
width of $\Upsilon(9460)$ in Ref.~\cite{hall} favors  a smaller value 4.1~GeV than that in
Ref.~\cite{2loopmb}. If we use a relatively 
small value $\overline{m}_b=$4.1GeV, the upper bound is reduced to be
0.65\%.  On the other hand, the Heavy Quark Effective Theory gives a 
lower bound on $\overline{m}_b$
($\overline{m}_b>4.26$GeV in Ref.~\cite{ligeti} and $\overline{m}_b>4.2$GeV in 
Ref.~\cite{luke}). If $\overline{m}_b$=4.5GeV, the upper bound weakens to 0.81\%. 

In summary, we pointed out that there is a $\tan^2 \beta$ enhancement in 
$\Gamma (Z \rightarrow \tau^+ \tau^-)$ induced by the neutral Higgs boson loops 
in the Type-II 2HDM,  whenever there is a similar 
enhancement to $R_b$.  We found that the current experimental upper bound 
$Y_{\tau}$ $< 1.0057$ at the 95\% confidence level places an upper 
bound $\Delta R_{b}/R_{b}$ $< 0.73\%$ (for $\overline{m}_b = 4.3$~GeV
and $\alpha_s (m_Z) = 0.118$), which can be compared to the 1996 
world average: $\Delta R_{b}/R_{b}$ = $0.97\% \pm 0.51\%$.


\section*{Acknowledgments}

S.K. would like to thank Y.~Okada, M.~Tanabashi, M.~Nojiri, S.~Kanemura, and N.~Tsuda
for helpful discussions. The work of J.H. was supported by the U.S. 
Department of Energy under Contract DE-FG02-94ER40823.  The work of 
H.M. was supported by the U.S. Department of Energy under Contract
DE-AC03-76SF00098 and in part by the National Science Foundation 
under grant PHY-95-14797.  H.M. was also supported by the Alfred P. Sloan 
Foundation.

\newpage
%
%
\newcommand{\Journal}[4]{{\sl #1}\/ {\bf #2} {(#3)} {#4}}
\newcommand{\APJ}{Ap. J.}
\newcommand{\CJP}{Can. J. Phys.}
\newcommand{\NC}{Nuovo Pimento}
\newcommand{\NP}{Nucl. Phys.}
\newcommand{\PL}{Phys. Lett.}
\newcommand{\PR}{Phys. Rev.}
\newcommand{\PRep}{Phys. Rep.}
\newcommand{\PRL}{Phys. Rev. Lett.}
\newcommand{\PTP}{Prog. Theor. Phys.}
\newcommand{\SJNP}{Sov. J. Nucl. Phys.}
\newcommand{\ZP}{Z. Phys.}

\end{document}